\documentstyle[12pt,psfig]{article}

\def\appendix{\par
 \setcounter{section}{0}
 \setcounter{subsection}{0}
 \def\thesection{Appendix \Alph{section}}}

\newcommand{\gone}[1]{}

\def\TL{\hfil$\displaystyle{##}$}
\def\TR{$\displaystyle{{}##}$\hfil}
\def\TC{\hfil$\displaystyle{##}$\hfil}

\def\seqalign#1#2{\vcenter{\openup1\jot
  \halign{\strut #1\cr #2 \cr}}}

\def\comment#1{}
\def\fixit#1{}

\def\tf#1#2{{\textstyle{#1 \over #2}}}

\def\mop#1{\mathop{\rm #1}\nolimits}

\def\Disc{\mop{Disc}}

\def\overleftrightarrow#1{\vbox{\ialign{##\crcr
     $\leftrightarrow$\crcr\noalign{\kern-0pt\nointerlineskip}
     $\hfil\displaystyle{#1}\hfil$\crcr}}}

\def\lsim{\mathrel{\mathstrut\smash{\ooalign{\raise2.5pt\hbox{$<$}\cr\lower2.5pt\hbox{$\sim$}}}}}
\def\gsim{\mathrel{\mathstrut\smash{\ooalign{\raise2.5pt\hbox{$>$}\cr\lower2.5pt\hbox{$\sim$}}}}}
\def\sqr#1#2{{\vcenter{\vbox{\hrule height.#2pt
         \hbox{\vrule width.#2pt height#1pt \kern#1pt
            \vrule width.#2pt}
         \hrule height.#2pt}}}}

\def\href#1#2{#2}  

\def\lbldef#1#2{\expandafter\gdef\csname #1\endcsname {#2}}
\def\eqn#1#2{\lbldef{#1}{(\ref{#1})}%
\begin{equation} #2 \label{#1} \end{equation}}
\def\eqalign#1{\vcenter{\openup1\jot
    \halign{\strut\span\TL & \span\TR\cr #1 \cr
   }}}
\def\eno#1{(\ref{#1})}
\def\O{{\cal O}}

\begin{document}
%--------+---------+---------+---------+---------+---------+---------+
%Title page
\begin{titlepage}

\begin{flushright}
PUPT-1791 \\
NSF-ITP-98-058\\
hep-th/9805140
\end{flushright}
%\vspace{12 mm}
\vfil\vfil

\begin{center}

{\Large {\bf Exact absorption probabilities for the D3-brane}}

\vfil

Steven S.~Gubser$^a$ and Akikazu Hashimoto$^b$

\vfil

$^a$Joseph Henry Laboratories\\
 Princeton University\\
 Princeton, New Jersey 08544\\

\vfil

$^b$Institute for Theoretical Physics\\
 University of California\\
 Santa Barbara, CA  93106

\vfil

\end{center}

%\vspace{10mm}

\vspace{5mm}

\begin{abstract} 
\noindent 
 We consider a minimal scalar in the presence of a three-brane in ten
dimensions.  The linearized equation of motion, which is just the wave
equation in the three-brane metric, can be solved in terms of
associated Mathieu functions.  An exact expression for the reflection
and absorption probabilities can be obtained in terms of the
characteristic exponent of Mathieu's equation.  We describe an
algorithm for obtaining the low-energy behavior as a series expansion,
and discuss the implications for the world-volume theory of D3-branes.
\end{abstract}

%\vspace{20mm}
\vfil\vfil\vfil
\begin{flushleft}
May 1998
\end{flushleft}
\end{titlepage}

\newpage
\renewcommand{\baselinestretch}{1.05}  %Line spacing

%--------+---------+---------+---------+---------+---------+---------+
%Body
\section{Introduction}
\label{Introduction}

One of the intriguing aspects of Ramond-Ramond solitons in string
theory is the existence of two alternative descriptions, one in terms
of supergravity \cite{HS91} and the other in terms of Dirichlet
branes (D-branes) \cite{JP}.  The description in terms of D-branes is
essentially perturbative in nature: each boundary picks up a factor of
$gN$, which is the square of the open string coupling times a
Chan-Paton factor.  As realized in \cite{Witten}, the low-energy
dynamics of $N$ coincident D-branes is dictated by maximally
supersymmetric gauge theory with gauge group $U(N)$, and $gN$ is
recognized as the 't Hooft parameter.

That the gauge theory and supergravity descriptions should be related
was implicit in much early work on absorption and Hawking emission
(see for example \cite{dmOne,Kleb97}).  A precise formulation of the
duality between the two descriptions was conjectured recently in
\cite{JuanAds97} by taking the so-called ``decoupling limit.''  The
simplest example comes from considering D3-branes in the type IIB
theory.  In the decoupling limit one obtains a duality between ${\cal
N}=4$ supersymmetric Yang-Mills theory in four dimensions and string
theory on the near horizon $AdS_5 \times S^5$ background
\cite{JuanAds97}.  The $AdS_5$ and the $S^5$ have the same radius of
curvature $R$, where $R^4 = 4 \pi gN \alpha'^2$.

It is difficult to find non-trivial checks of the duality because it
relates two things that are rather poorly understood away from certain
limits.  On the $AdS$ side, it is widely felt that the supergravity
description should be capable of being elevated to a full closed
string theory description, similar to non-linear sigma models; but it
is not understood how to include Ramond-Ramond backgrounds in a
non-linear sigma model.\footnote{See however
\cite{Russo:1998xv,Metsaev:1998it} for interesting recent work on
including Ramond-Ramond fields in a world-sheet formulation.}  We
must for the present content ourselves with the supergravity limit.
The validity of this limit relies on having a large number $N$ of
coincident branes, with a small closed string coupling $g$, but large
$gN$.  Large $gN$ is exactly where the gauge theory is difficult to
deal with: after 't~Hooft scaling, the Feynman rules associate a
factor $gN$ with each vertex, so for generic amplitudes one must
consider large graphs.

How then can we study the relation between the two dual descriptions
concretely?  Aside from the calculation of entropy \cite{gkp96}, one
of the simplest quantities that can be computed on both side of the
correspondence is the absorption cross-section of scalar fields
incident on the branes.  Suppression of stringy correction on the
supergravity side relies on having $\omega \sqrt{\alpha'} \ll 1 $ and
$\sqrt{\alpha'}/R \ll 1$; but $\omega R$ can be arbitrary, suggesting
the existence of a double scaling limit \cite{Kleb97,GKT97}. Indeed,
the wave equation for the fields propagating in the supergravity
background of branes depend on $gN$ only in the combination $\omega
R$.  Remarkably, the leading order behavior in small $\omega R$ of the
semi-classical cross-section is reproduced by a tree level gauge
theory calculation (leading order in $gN$) \cite{Kleb97,GKT97}.  The
relevant gauge theory amplitude apparently suffers no radiative
corrections.  An argument for why this is so was advanced in
\cite{GKSchwing} for graviton absorption, and other examples have
emerged in \cite{MIT,Chalmers}.

A natural question which arises at this point is whether this pattern
persists to higher order in $\omega R$ \cite{Das97}.  In order to
address this question, one must examine higher order corrections in
both D-brane and supergravity computations. On the supergravity side,
a first step in this direction was taken in \cite{GHKK98} were terms
subleading by order $(\omega R)^4$ was examined. The coefficient of
the $(\omega R)^4$ correction turns out to have a piece which depends
logarithmically in $\omega R$:
  \eqn{SigmaForm}{
   \sigma = {\kappa^2 N^2 \omega^3 \over 32 \pi} \left[1 + 
    c'_1 (\omega R)^4 \log \omega R + c_1 (\omega R)^4 + 
     {\cal O}((\omega R)^8)\right]
  }
and the numerical value of $c_1'$ was found to be $-1/6$.

The goal of this paper is to describe an algorithm for computing the
absorption cross-section as a power series expansion in $\omega R$
to all orders.  The absorption cross-section is determined by
comparing the flux of incident partial waves at asymptotic region and
the near horizon region. We are therefore interested in finding the
solution to the wave equation of scalar fields in the background of
the D3-brane metric. It turns out that the wave equation in question
is equivalent to Mathieu's modified differential equation\footnote{The
usual form of Mathieu's equation is obtained from \eno{AMat} via the
replacement $z \to iz$.}
\eqn{AMat}{
\left[ {\partial^2 \over \partial z^2} + 2 q \cosh 2z - a \right] \psi(z) = 0
}
 under appropriate change of variables and field redefinitions.  The
exact solution of Mathieu's modified differential equation is known in
the form of power series expansion with respect to $q$.  From this, we
can read off the absorption cross-section. For reviews of Mathieu
functions see \cite{hmf,GR,whit,bate,lach}.  In view of the relative
obscurity of these functions, most of the relevant details will be
included in our exposition.

First, let us see how Mathieu's modified equation arises from the wave
equation of scalar fields. The supergravity background for the
D3-brane has the simple form \cite{HS91}
$$
ds^2 = H^{-1/2} (-dt^2 + dx_\parallel^2) + H^{1/2} dx_\perp^2 
$$
as well as some RR 4-form background, where
$$H = 1 + {R^4 \over r^4} \ , \qquad 
  R^4 = 4 \pi g N \alpha'^2 = {N \kappa \over 2 \pi^{5/2}} \ .$$
For scalar fields decoupled from the RR 4-form (the example we will
always have in mind is the dilaton), the equation of motion
is simply
$$\frac{1}{\sqrt{g}} \partial_\mu \sqrt{g} 
  g^{\mu \nu} \partial_\nu \phi = 0 \ .$$
The radial wave equation for the $l$-th partial wave of energy
$\omega$ which follows from this equation is
\begin{equation}
\left[{\partial^2 \over \partial r^2} + {5 \over r} {\partial
\over \partial r} - {l (l+4) \over r^2} + \omega^2 \left(1+ {R^4
\over r^4} \right) \right] \phi^{(l)}(r) = 0 . 
\label{maineq}
\end{equation}
In order to relate (\ref{maineq}) to Mathieu's equation alluded to
earlier, one performs the following change of variables:\comment{I
sent $z \to -z$ relative to our earlier conventions: this is more
consistent with using $H^{(1)}(\nu,z)$ as the tunneling solutions.}
$$r = R e^{-z}, \qquad \phi(r) = e^{2z} \psi(z).$$
In terms of these new variables, equation (\ref{maineq}) reads
\begin{equation}
\left[ {\partial^2 \over \partial z^2} + 
 2 (\omega R)^2 \cosh 2z -  (l+2)^2 \right] \psi(z) = 0.
\label{maineq2}
\end{equation}
 which is precisely of the form (\ref{AMat}) for $q = (\omega R)^2$ and $a
= (l+2)^2$.  Note that we have reduced the problem of particle
absorption by three-branes to the computation of the tunneling
$S$-matrix for a one-dimensional Schrodinger equation.\footnote{As an
aside we note that the equations of motion for supergravity fields
other than minimal scalars generically do not lead to the Mathieu
equation.  For example, the fixed scalar considered in \cite{GHKK98}
experiences a ``transmutation of angular momentum,'' in the sense that
the low-energy radial function at infinity and near the horizon are
Bessel functions of different orders.  To put it differently, the
potential function in the Schrodinger operator is asymmetric.}

The rest of the paper is organized as follows.  In
section~\ref{Mathieu} we present the method for obtaining the
absorption probability from Mathieu's equation.  This method will be
of primary interest to the mathematically oriented reader, but those
concerned with the string theory implications may wish to skip
directly to the final answer, \eno{PForm}.  Section~\ref{World} is
concerned with the world-volume interpretation of this probability.
Section~\ref{Discuss} concludes with a brief discussion.  The appendix
includes some formulas judged too cumbersome to include in the main
text.

\section{Cross-sections from Mathieu functions}
\label{Mathieu}

Mathieu functions arise in the study of a variety of physical
problems: for example, the solution of the flat-space Laplace equation
in elliptical coordinates; Bloch waves for the potential $\cos 2x$;
the Faraday instability; classical motion of a driven pendulum; the
sine-Gordon model \cite{neu}; and, in the present context, as
tunneling wavefunctions in the potential $-\cosh 2z$.  Our analysis is
an extension of \cite{doug}, and our conventions will be a hybrid of
those of \cite{hmf} and \cite{doug}.

The so-called Floquet solutions of \AMat\ can be expressed in the form
  \eqn{Bloch}{
   J(\nu,z) = \sum_{n=-\infty}^\infty 
    \phi\left( n + \tf{1}{2} \nu \right) e^{(2n+\nu) z} \ .
  } 
 These solutions are analogous to Bloch waves because of the property
  \eqn{ShiftOp}{
   J(\nu,z+i\pi) = e^{i\pi\nu} J(\nu,z) \ .
  }
The quantity $\nu$ is termed the Floquet exponent and is determined in
terms of $a$ and $q$.  Clearly, $J(\nu,-z)$ is also a solution of
\AMat .  Since $J(\nu,-z)$ acquires a phase $e^{-i \pi \nu}$ under $z
\rightarrow z + i \pi$, $J(\nu,-z)$ is also a Floquet solution with
exponent $-\nu$. It follows that there is a proportionality relation
  \eqn{FloqRel}{
   J(-\nu,z) \propto J(\nu,-z) \ ,
  }
which will become useful in the later discussions.  

It is
straightforward to see that \Bloch\ solves \AMat\ if \eqn{PhiEqn}{
\phi(z+1) + \phi(z-1) = {z^2-r^2 \over \lambda^2} \phi(z) } where we
have defined $r = \tf{1}{2} \sqrt{a}$ and $\lambda = \tf{1}{2}
\sqrt{q}$.  A meromorphic function $\phi$ was found in \cite{doug}
which satisfies the recursion relation \PhiEqn\ and in addition has
the property $\phi\to 0$ as $\Re z\to\infty$.\comment{Don't change
this!  The properties of $\phi$ as $\Re z\to -\infty$ are complicated,
and only the relation \eno{recur} keeps the actual coefficients in
\Bloch\ from blowing up.}  Explicitly, \eqn{DougPhi}{\eqalign{ \phi(z)
&= {\lambda^{2z} \over \Gamma(z+r+1) \Gamma(z-r+1)} v(z) \cr v(z) &=
\sum_{n=0}^\infty (-1)^n \lambda^{4n} A_z^{(n)} \cr A_z^{(0)} &= 1 \cr
A_z^{(q)} &= \sum_{p_1=0}^\infty \sum_{p_2=2}^\infty \ldots
\sum_{p_q=2}^\infty a_{z+p_1} a_{z+p_1+p_2} \cdots
a_{z+p_1+\ldots+p_q} \cr a_z &= {1 \over (z+r+1)(z+r+2)(z-r+1)(z-r+2)}
\ .  }}

The value of $\nu=2\mu$ is determined by relation \FloqRel, which
implies
\eqn{recur}{{\phi(\mu) \over \phi(\mu-1)} \times 
  {\phi(-\mu+1) \over \phi(-\mu)}=1 \ .}
The recursion relation \PhiEqn\ can be written in the form
$$  V(z)=  {\phi(z+1) \over \phi(z)} + {\phi(z-1) \over \phi(z)} = G_{z+1} + {1 \over G_z} $$
where we have defined $V(z) = {(z^2-r^2) / \lambda^2} $ and $G_z =
\phi(z)/\phi(z-1)$. Then, we can express the first factor  of \recur\
as a continued fraction:
$${\phi(\mu) \over \phi(\mu-1)} = G_\mu = 
   {1 \over V(\mu) - G_{\mu + 1}} = 
   {1 \over V(\mu) - } {1 \over V(\mu+1) - \ldots} \ .$$
Similarly, the recursion relation \PhiEqn\ can be written in yet another form
$$  V(z)=  {\phi(-z+1) \over \phi(-z)} + {\phi(-z-1) \over \phi(-z)} = H_{z-1} + {1 \over H_z} $$
where this time we have defined $H_z = \phi(-z)/\phi(-z-1)$. Now we
can also express the second factor  of \recur\ as a continued fraction:
$${\phi(-\mu+1) \over \phi(-\mu)} = H_{\mu-1} = 
   {1 \over V(\mu-1) - H_{\mu -2}} = 
   {1 \over V(\mu-1) - } {1 \over V(\mu-2) - \ldots} \ .$$
It is now straightforward to solve for $\mu$ order by order in $\lambda$. We simply substitute  the ansatz
$$\nu = \nu_0 + \nu_1 \lambda^4 + \nu_2 \lambda^8 + \ldots$$
into \recur\ expressed in terms of the continued fractions.  If we are
only interested in the value of $\nu$ to some finite order in
$\lambda$, we can truncate the continued fraction by finite iteration.
In equation (\ref{nuexp}) of the appendix we give the first few terms
of the series for the partial waves $l=0$, $l=1$, and $l=2$.

There is a remarkable resummation of the Bloch wave expansion \Bloch\
in terms of Bessel functions:\footnote{We use a notational convention
where $J_\nu(z)$ with subscript $\nu$ denote Bessel functions whereas
$J(\nu,z)$ with argument $\nu$ denote solutions to Mathieu's equation
\AMat.}
  \eqn{DoubExp}{
   J(\nu,z) = \sum_{n=-\infty}^\infty 
    {\phi\left( n + \tf{1}{2} \nu \right) \over 
     \phi(\nu/2)} J_n(\sqrt{q} e^{-z})
     J_{n+\nu}(\sqrt{q} e^z) \ .
  } 
The expansion \DoubExp\ is uniformly convergent
everywhere\comment{``has the advantage'' wasn't really right, since
the original expansion also converges everywhere.-- I agree. } and is
convenient for extracting the asymptotic behavior for large $|z|$
\cite{doug,bate}.  For $\nu \notin {\bf Z}$, the Floquet solutions
$J(\pm\nu,z)$ are independent.  It is useful, however, to consider
other linear combinations, $N(\nu,z)$, $H^{(1)}(\nu,z)$, and
$H^{(2)}(\nu,z)$, in analogy with Bessel functions:
  \eqn{OtherMat}{\eqalign{
   N(\nu,z) &= {\cos \pi\nu \, J(\nu,z) - J(-\nu,z) \over \sin \pi\nu}  \cr
   H^{(1)}(\nu,z) &= J(\nu,z) + i N(\nu,z) = 
    {J(-\nu,z) - e^{-i\pi\nu} J(\nu,z) \over i \sin \pi\nu}  \cr
   H^{(2)}(\nu,z) &= J(\nu,z) - i N(\nu,z) = 
    {J(-\nu,z) - e^{i\pi\nu} J(\nu,z) \over -i \sin \pi\nu}\ .  }}
Some useful  relations among the various solutions are
  \eqn{OtherMatRel}{\eqalign{
   J(\nu,z) &= {H^{(1)}(\nu,z) + H^{(2)}(\nu,z) \over 2}  \cr
   J(-\nu,z) &= {e^{i\pi\nu} H^{(1)}(\nu,z) + 
     e^{-i\pi\nu} H^{(2)}(\nu,z) \over 2} \ .
  }}
Using \recur\ and the standard relation $J_{-n} = (-1)^n J_n$, it is
straightforward to show that solutions \OtherMat\ also admit
expansions in terms of Bessel functions, generalizing \DoubExp:
  \eqn{AnyExp}{
   Z^{(j)}(\nu,z) = \sum_{n=-\infty}^\infty 
    {\phi\left( n + \tf{1}{2} \nu \right) \over 
     \phi(\nu/2)} J_n(\sqrt{q} e^{-z})
     Z^{(j)}_{n+\nu}(\sqrt{q} e^z) \ .
  }
Here, $Z^{(j)}$ runs over $J$, $N$, $H^{(1)}$, and $H^{(2)}$.  These
solutions are termed associated Mathieu functions of the first,
second, third, and fourth kinds.\footnote{We emphasize, however, that
of these only $J(\nu,z)$ is a Floquet solution.}  We will primarily be
interested in the third kind, since that is the one which describes
tunneling from asymptotic infinity into the three-brane.

The asymptotic behavior for $\Re z \to \infty$ is manifest from the
expansion \AnyExp:
  \eqn{BigBehave}{
   Z^{(j)}(\nu,z) \to Z^{(j)}_\nu(\sqrt{q} e^z) 
    \qquad\hbox{as $\Re z \to \infty$.}
  }
 The behavior for $\Re z \to -\infty$ is more difficult to decipher.
The first step is to use \recur\ to show that the constant of
proportionality in \FloqRel\ is precisely $\phi(-\nu/2)/\phi(\nu/2)$:
  \eqn{FlipRel}{
   J(-\nu,z) = {\phi(-\nu/2) \over \phi(\nu/2)} J(\nu,-z) \ .
  }
 It is useful at this point to introduce the two quantities
  \eqn{EtaChi}{
   \eta = e^{i\pi\nu} \qquad \chi = {\phi(-\nu/2) \over \phi(\nu/2)} \ .
  }
Now the behavior of $H^{(1)}(z)$ as $\Re z \to -\infty$ can be
investigated by using \OtherMat, \OtherMatRel\ and {\FlipRel}:
  \eqn{HMinus}{
   H^{(1)}(\nu,z) = {1 \over 2i\sin\pi\nu} \left[ 
        \left( \chi - {1 \over \chi} \right) H^{(1)}(\nu,-z) + 
        \left( \chi - {e^{-2i\pi\nu} \over \chi} \right)
         H^{(2)}(\nu,-z) \right] \ .
  }
Recalling the asymptotics
  \eqn{BesselAsympt}{
   \left. \eqalign{
    H_\nu^{(1)}(\xi) &\to
     \sqrt{2 \over \pi\xi} 
      e^{i \left(\xi - {\pi \over 2} \nu - {\pi \over 4}\right)}  \cr
    H_\nu^{(2)}(\xi) &\to
     \sqrt{2 \over \pi\xi} 
      e^{-i \left(\xi - {\pi \over 2} \nu - {\pi \over 4}\right)}  
   }\right\} \qquad\hbox{as $\Re\xi \to \infty$}
  }
 we obtain
  \eqn{BothSides}{
   \sqrt{\eta} \left( \eta - {1 \over \eta} \right) H^{(1)}(\nu,z) 
    \to \left\{
     \eqalign{
      & \left( \eta - {1 \over \eta} \right)
         \sqrt{2 \over \pi \sqrt{q} e^z} 
         e^{i \left( \sqrt{q} e^z - {\pi \over 4} \right)} 
        \quad\hbox{for $\Re z \to \infty$}  \cr
      & \left( \chi - {1 \over \chi} \right) 
         \sqrt{2 \over \pi \sqrt{q} e^{-z}} 
         e^{i \left( \sqrt{q} e^{-z} - {\pi \over 4} \right)}  \cr
      &\qquad + \left( \eta \chi - 
        {1 \over \eta \chi} \right)
         \sqrt{2 \over \pi \sqrt{q} e^{-z}} 
         e^{-i \left( \sqrt{q} e^{-z} - {\pi \over 4} \right)}
         \quad\hbox{for $\Re z \to -\infty$} \ .
     } \right. 
  }
 From \BothSides\ we read off the amplitudes $A = \chi - {1 \over
\chi}$, $B = \chi \eta - {1 \over \chi \eta}$, and $C = \eta - {1
\over \eta}$ for the reflected, incident, and transmitted waves,
respectively.  

A consistency check on \BothSides\ is the unitarity relation, $|B|^2 =
|A|^2 + |C|^2$.  One way to prove this relation is to send $z \to z +
i\pi/2$ so that the $-\cosh$ potential is inverted to $+\cosh$.
Clearly there are wavefunctions in this potential which are everywhere
real and exponentially decaying on one side (but not the other unless
$a$ is an eigen-energy).\footnote{Incidentally, \BothSides\ provides
an implicit equation for the eigen-energies of the $+\cosh$ potential:
namely $\chi = \pm 1$ for even/odd wavefunctions.}  In fact,
$H^{(1)}(z+i\pi/2)$ is just such a solution, up to a constant overall
phase.  Hence $A/C$ is pure imaginary.  Now, $2\cos\pi\nu = \eta + {1
\over \eta}$ is always real for real $q$ (a consequence of Hill's
equation).  Hence $\eta$ is always either real or of unit modulus.
The statement that $A/C$ is imaginary means that the same is true of
$\chi$, and moreover $\chi$ is real when $\eta$ is of unit modulus and
vice versa.  The verification of unitarity,
  \eqn{Unitary}{
   \left| \eta \chi - {1 \over \eta \chi} \right|^2 = 
    \left| \eta - {1 \over \eta} \right|^2 + 
    \left| \chi - {1 \over \chi} \right|^2 \ ,
  }
is now straightforward.  It proves easiest in practice to compute the
absorption probability from
  \eqn{AbsProb}{
   P = {\left| \eta - {1 \over \eta} \right|^2 \over
    \left| \eta - {1 \over \eta} \right|^2 + 
    \left| \chi - {1 \over \chi} \right|^2} \ ,
  }
but of course there are several equivalent alternative forms.

Following the methods of \cite{doug}, it is straightforward though
tedious to obtain a series expansion of $\chi$ in $q$.  The first
observation is that any formal sum
  \eqn{DougSum}{
   A_q = \sum_{p_1=-\infty}^\infty \sum_{p_2=2}^\infty \ldots
    \sum_{p_q=2}^\infty t_{p_1} t_{p_1+p_2} \cdots t_{p_1+\ldots+p_q}
     \ , 
  }
where the $t_n$ are regarded as independent variables, can be written
in terms of products of single sums of products of the $t_n$.  A
recursion relation is derived in \cite{doug} to demonstrate this fact:
  \eqn{RecurA}{\eqalign{
   q A_q &= \sum_{n=-\infty}^\infty 
     \left( t_n {\partial A_q \over \partial t_n} \right)  \cr
    &= \sum_{n=-\infty}^\infty \left[ t_n -
     t_n (t_{n-1} + t_n + t_{n+1}) {\partial \over \partial t_n} +
     t_{n-1} t_n t_{n+1} 
     {\partial^2 \over \partial t_{n-1} \partial t_{n+1}} \right]
     A_{q-1} \ .
  }}
 Let us introduce the notation
  \eqn{SNote}{
   S[\alpha_0,\alpha_1,\ldots,\alpha_k] = 
    \sum_{n=-\infty}^\infty \prod_{j=0}^k t_{n+j}^{\alpha_j} \ ,
  }
 where the $\alpha_j$ are natural numbers with $\alpha_0$ and
$\alpha_k$ nonzero.\footnote{Note that only questions of convergence
stand in the way of extending the following discussion to arbitrary
real sequences $\left\{ \alpha_j \right\}_{j=-\infty}^\infty$ modulo
the equivalence relation $\left\{ \alpha_j \right\}_{j=-\infty}^\infty
\sim \left\{ \alpha_{j+k} \right\}_{j=-\infty}^\infty$ for integer
$k$.}  Then the map $H: A_{q-1} \to q A_q$ defined in \RecurA\ can be
viewed formally as a linear operator on the infinite-dimensional
vector space whose basis is $1$ together with all possible products of
the $S[\alpha_0,\alpha_1,\ldots,\alpha_k]$.\footnote{This space is
reminiscent of the loop spaces encountered, for instance, in the $c=0$
matrix model \cite{ish,gSeniorThesis}.  In this analogy, $H$ plays the
role of the Fokker-Planck Hamiltonian.}  We have
  \eqn{GotA}{
   A_q = {1 \over q!} H^q(1)
  }
where $H^q(1)$ is the operator $H$ acting $q$ times on unity.
Amusingly, the problem of computing the generalization of the function
$v$ in \DougPhi\ to arbitrary $t_n$ at finite $\lambda$ is formally
identical to Euclidean evolution by the Hamiltonian $H$:
  \eqn{vEvolve}{
   v \equiv \sum_{q=0}^\infty (-\lambda^4)^q A_q
     = e^{-\lambda^4 H}(1) \ .
  }%
\begin{sloppypar}
In equation (\ref{SeveralA}) of the appendix we write out the first four
$A_q$ in terms of the $S[\alpha_0,\alpha_1,\ldots,\alpha_k]$.
\end{sloppypar}

Now let us specialize to $A_q = A_z^{(q)}$ by setting
  \eqn{SpecialT}{
   t_n = \left\{ \eqalign{a_{z+n} &\quad\hbox{for $n \geq 0$}  \cr
                          0 &\quad\hbox{otherwise.} }\right.
  }
 The sums $S[\alpha_0,\alpha_1,\ldots,\alpha_q]$ then have the general
form $\sum_{n=0}^\infty {1 \over f(z+n)}$ where $f(z)$ is a polynomial
of degree $4 \sum_{i=0}^q \alpha_i$.  Such sums can be performed
explicitly in terms of the function $\psi(z) = \Gamma'(z)/\Gamma(z)$
and its derivatives.  The first step is to make a partial fraction
decomposition:
  \eqn{PFrac}{
   {1 \over f(z)} = \sum_{f(y)=0} \sum_{\ell=1}^\infty 
    {c^{(\ell)}_y \over (z-y)^\ell} \ .
  }
 The first sum is over the roots of $f(z)$.  For a root $y$ of
multiplicity $k$, only the first $k$ of the constants $c^{(\ell)}_y$
can be nonzero.  Each term in the partial fraction decomposition makes
a contribution to the sum over $n$ which can be read off from 
  \eqn{PsiDefs}{\eqalign{
   \psi(z) &= -{\bf C} - \sum_{n=0}^\infty \left[
    {1 \over z+n} - {1 \over n+1} \right]  \cr
   \psi^{(k)}(z) &= (-1)^{k+1} k! \sum_{n=0}^\infty 
    {1 \over (z+n)^{k+1}} \ .
  }}
 where ${\bf C} = \log\gamma \approx 0.5772$ is Euler's
constant. This leads to the sum
  \eqn{FinalSum}{
   \sum_{n=0}^\infty {1 \over f(z+n)} = 
    \sum_{f(y)=0} \sum_{\ell=1}^\infty c^{(\ell)}_y 
     {(-1)^\ell \over (\ell-1)!} \psi^{(\ell-1)}(z-y) \ .
  }
 The coefficients $c_y^{(1)}$ satisfy the relation
  \eqn{AllResidues}{
   \sum_{f(y)=0} c^{(1)}_y = {1 \over 2\pi i} \oint_\gamma 
    {dz \over f(z)} = 0 \ ,
  }
where $\gamma$ is a contour that encloses all the roots of $f(z)$,
ensuring that the divergences from the various $1/(z-y)$ terms in the
partial fraction decomposition cancel.  In effect this allows us to
use the second line of \PsiDefs\ even at $k=0$.  Explicit expressions
for the first few $S[\{\alpha_i\}]$'s are included in equation
(\ref{ExplicitS}) of the appendix.

To complete the task of computing the absorption cross-section, we
need to determine the value of $\chi = \phi(-\nu/2) /
\phi(\nu/2)$. All that remains to be done now is to substitute the
expansion for $\nu$ given in (\ref{nuexp}) into \DougPhi\ and collect
terms of given order in $\lambda$.  Because $\nu$ is an integer plus
powers of $\lambda$, the $\psi$ functions can all be Taylor expanded
around integers or half-integers.  To simplify the final expressions,
it is useful to recall the relation of $\psi$ to the Riemann zeta
function $\zeta(s)$ and its generalizations $\zeta(s,z)$:
  \eqn{PsiValues}{\seqalign{\span\TC}{
   \psi(1) = -{\bf C} \qquad
   \psi^{(k)}(z) = (-1)^{k+1} k! \zeta(k+1,z)  \cr
   \zeta(s,z+1) = \zeta(s,z) - {1 \over z^s}\cr
   \zeta(s,1) = \zeta(s) \qquad 
   \zeta\left( s,\tf{1}{2} \right) = (2^s-1) \zeta(s) \ .
  }}
The final expressions for the absorption probability of the $l$-th
partial wave have the form
\begin{equation}
   P_l = {4 \pi^2 \over (l+1)!^4 (l+2)^2} 
    \left( \omega R/2 \right)^{8+4l}
    \sum_{n=0}^\infty \sum_{k=0}^n b_{n,k} (\omega R)^{4n} 
     \left( \log \omega \bar{R} \right)^k \ ,
\label{PForm}
\end{equation}
where $\bar{R} = e^{\bf C} R / 2$.  The overall normalization have
been chosen so that $b_{0,0}=1$.  We have computed the values of the
first few $b_{n,k}$'s for $l=0$, $l=1$, and $l=2$ which we summarize
in Table \ref{table1}. We find that $b_{n,k}$ is rational for $n-k<2$,
whereas for $n-k \geq 2$ it is a linear combination of $\zeta(2),
\zeta(3), \ldots, \zeta(n-k)$ with rational coefficients.

\begin{table}[t]
\newcommand{\spac}{\raise-1.3ex \hbox{\rule{0ex}{4.0ex}}}
\begin{tabular}{|l||r|r|r|} \hline
\spac & $l=0$ & $l=1$ & $l=2$ \\ \hline \hline
$ \spac \! b_{1,1} \!$ & $-{\frac{1}{6}}$& $-{\frac{1}{24}}$& $-{\frac{1}{60}}
  $ \\ \hline
$ \spac \!   b_{1,0} \!$ & ${\frac{7}{72}}$& ${\frac{53}{1152}}$& $
  {\frac{19}{800}}$ \\ \hline \hline
$ \spac \! b_{2,2} \!$ & ${\frac{17}{576}}$& ${\frac{1}{1152}}$& $
  {\frac{1}{7200}}$ \\ \hline
$ \spac \! b_{2,1} \!$ & $-{\frac{161}{4608}}$& $-{\frac{757}{276480}}$& $
  -{\frac{821}{1728000}}$ \\ \hline
$ \spac \! b_{2,0} \!$ & ${\frac{5561}{663552}} - {\frac{11\zeta(2)}{576}}$& $
  {\frac{261343}{132710400}} - {\frac{\zeta(2)}{4608}}$& $
  {\frac{44071}{103680000}} - {\frac{\zeta(2)}{28800}}$ \\ \hline \hline
$ \spac \! b_{3,3} \!$ & $-{\frac{11}{2592}}$& $-{\frac{1}{82944}}$& $
  -{\frac{1}{1296000}}$ \\ \hline
$ \spac \! b_{3,2} \!$ & ${\frac{623}{82944}}$& ${\frac{7}{69120}}$& $
  {\frac{479}{103680000}}$ \\ \hline
$ \spac \! b_{3,1} \!$ & $-{\frac{39037}{9953280}} + {\frac{49\zeta(2)}{6912}}
  $& $-{\frac{554911}{3185049600}} + {\frac{\zeta(2)}{110592}}$& $
  -{\frac{1731599}{174182400000}} + {\frac{\zeta(2)}{1728000}}$ \\ \hline
$ \spac \!  b_{3,0}\! $ &  $\!\! {\frac{1093099}{2388787200}} \! - \!   {\frac{1379\zeta(2)}{331776}} \! + \! {\frac{5\zeta(3)}{41472}}
 \!$  &  $ \!\!
  {\frac{65129557}{764411904000}} \! - \! {\frac{101\zeta(2)}{2211840}} \! - \!
   {\frac{\zeta(3)}{663552}}\! $  &  $\!\!
 {\frac{1148018521}{167215104000000}}  \! - \!  {\frac{479\zeta(2)}{414720000}} \!- \!
   {\frac{\zeta(3)}{10368000}} \! $  \\ \hline
\end{tabular} 
\caption{Leading coefficients $b_{n,k}$ for the expansion with respect
to $\omega R$ for the absorption cross-section (\ref{PForm}) of $l=0$,
$l=1$, and $l=2$ partial waves.\label{table1}}
\end{table}

The absorption cross-section for the $l$-th partial wave can now be
computed from a version of the Optical Theorem:
  \eqn{OptSigma}{
   \sigma_l = {8 \pi^2 / 3 \over \omega^5} (l+1) (l+2)^2 (l+3) P_l \ .
  }
 The generalization of this formula to arbitrary dimensions was
derived in \cite{gpartial}.

\section{The world-volume dynamics}
\label{World}

Let us now consider the world-volume interpretation for the case where
the minimal scalar is the dilaton.  In the 't Hooft limit $g
\rightarrow 0$, $N\rightarrow \infty$ with $gN$ fixed, quantum
fluctuations of bulk fields decouple and the dynamics is strictly on
the brane world-volume.  The only sense in which bulk fields enter is
as a source of world-volume fluctuations in the form of a local
operator. The $s$-wave of the dilaton corresponds in the world-volume
theory to the operator $\O$ which slides the gauge coupling.  The
absorption probability $P_{l=0}$ then translates directly into the
discontinuity of the cut in the two-point function $\O$ through the
formula \cite{GKSchwing} \eqn{PPiForm}{\eqalign{ P_{l=0} &= {\pi^3
\omega^4 R^8 \over 8 i N^2} \Disc \Pi(p^2) \cr \Pi(p^2) &= \int d^4 x
\, e^{i p \cdot x} \Pi(x^2)}}
where
\begin{equation}
\Pi(x^2) = \langle \O(x) \O(0) \rangle \ .
\label{defineCorr}
\end{equation}
The dynamics of the world-volume theory at leading order in energy is
captured by its superconformal limit in the infrared. To higher order
in energy, however, one must account for the effect of irrelevant
perturbations which takes the theory away from the fixed point.  The
correlator $\langle \ldots \rangle$ is therefore taken with respect to
some quantum effective action which we will describe later in this
section.

In \PPiForm\ it should be noted that the discontinuity is taken across
the cut positioned along the positive real axis of the complex $s =
-p^2$ plane, evaluated at $s = \omega^2$.  Working backward, one can
read off $\Pi(x^2)$ from $P_{l=0}$, with the result
  \eqn{PiXForm}{
   \Pi(x^2) = {3 N^2 \over \pi^4 x^8} \sum_{n=0}^\infty
    \sum_{k=0}^n c_{n,k} \left( {R^2 \over x^2} \right)^{2n}
     \left( \log {R^2 \over x^2} \right)^k \ .
  }
 To obtain $P_{l=0}$ from \PiXForm\ we must specify a regularization
scheme for the Fourier integrals.  The minimal scheme, following
\cite{FreedDiff,AnsCentral}, is to analytically continue the formula
  \eqn{DiffScheme}{
   \int d^4 x {e^{i p \cdot x} \over x^{2h}} = 
    \pi^2 \left( {4 \over p^2} \right)^{2-h} 
    {\Gamma(2-h) \over \Gamma(h)}
  }
 beyond its radius of convergence $|h-1|<1$ to a meromorphic function
on the entire complex $h$ plane, and then read off the behavior near
the poles at positive integer $h$ by matching terms in the Taylor
expansions in $a$ of
  \eqn{DiscInts}{\eqalign{
   \int d^4 x {e^{i p \cdot x} (\mu x)^{2a} \over x^{2n}} &= 
    \pi^2 \left( {4 \over p^2} \right)^{2-n} 
          \left( {4 \mu^2 \over p^2} \right)^a
     {\Gamma(2-n+a) \over \Gamma(n-a)}  \cr
   \Disc \int d^4 x {e^{i p \cdot x} (\mu x)^{2a} \over x^{2n}} &= 
    -\left( 4 \over \omega^2 \right)^{2-n} 
     \left( 4 \mu^2 \over \omega^2 \right)^a
    {2 \pi^3 i \over \Gamma(n-a) \Gamma(n-a-1)} \ .
  }}
 For the expansions in $a$ one uses 
  \eqn{AExpands}{\eqalign{
   (\mu x)^{2a} &= 
    \sum_{n=0}^\infty {a^n \over n!} (\log \mu^2 x^2)^n  \cr
   \log \Gamma(1+a) &=
    \tf{1}{2} \log {\pi a \over \sin \pi a} - 
     {\bf C} a - \sum_{n=1}^\infty {a^{2n+1} \over 2n+1} \zeta(2n+1) \ .
  }}
 Upon setting the energy scale $\mu = 1/R$ one obtains the $c_{n,k}$
as numbers involving $\zeta(s)$ in the same way as the $b_{n,k}$:
explicitly,
  \eqn{QuoteCnk}{\seqalign{\span\TL&\span\TR&\qquad
                 \span\TL&\span\TR&\qquad
                 \span\TL&\span\TR}{
    c_{0,0} &= 1 & c_{1,1} &= -320 &  c_{2,2} &= 571200  \cr
            &    & c_{1,0} &= -1024 & c_{2,1} &= 4408560  \cr
            &    &         &        & c_{2,0} &=
     {14 \over 3} (1422697 - 12000 \pi^2) \ .
  }}
 One can formally define a dimension $\Delta$ for the operator $\O$ in
\eno{defineCorr} through a version of the Callan-Symanzik equation:
  \eqn{DefineDim}{
   \left[ x {\partial \over \partial x} + 2 \Delta \right]
    \Pi(x^2) = 0 \ .
  }
 For $R^4/x^4 \ll 1$, this results in a series of the same form as
\PiXForm: 
  \eqn{DeltaSeries}{\eqalign{
   \Delta &= \sum_{n=0}^\infty \sum_{k=0}^n \Delta_{n,k}
    \left( {R^2 \over x^2} \right)^{2n}
     \left( \log {R^2 \over x^2} \right)^k  \cr
    &= 4 - 64 {R^4 \over x^4} 
     \left( 37 + 10 \log {R^2 \over x^2} \right) + \ldots \ .
  }}

The challenge at this point is to reproduce \PiXForm\ and its
generalizations to higher partial waves through a quantum field theory
analysis.  As we mentioned earlier in this section, this requires a
knowledge of the world-volume dynamics beyond the superconformal limit
in the infrared.  In principle, this theory is well defined as a
low-energy effective action of the full string theory. At present,
however, no concrete formulation of this effective theory is
known. Therefore, instead of trying to reproduce \PiXForm, we can
attempt to learn about this effective theory from the data provided by
\PiXForm.

The leading term has precisely the form one expects in a conformal
theory.  The leading correction, ${R^4 \over x^4} \log {R^2 \over
x^2}$, has the form one would obtain by perturbing the conformal field
theory by a dimension eight operator.  It was speculated in
\cite{GHKK98} that this correction and perhaps the full semi-classical
cross-section would eventually find its world-volume explanation in
the non-abelian Dirac-Born-Infeld (DBI) action, with the symmetrized
trace prescription proposed in \cite{Tdbi} to pick out the leading
correction at dimension eight (${\rm Tr}[F^4]$), rather than dimension
six (${\rm Tr}[F^3])$ as one would expect from other prescriptions.

However, the DBI action arises from summing disc diagrams, so it defines a
classical field theory, and in no way captures the effect of a
resummation of infinite insertions of boundaries in the large $gN$
limit.  Furthermore, the non-renormalizability of the action makes it
impossible to proceed to the quantum theory from a knowledge of the
tree-level amplitudes alone, as was the standard strategy in deriving
low-energy renormalizable quantum field theories from string theory.
We require some further input from the string theory.

It was conjectured in \cite{gkPol,WitHolOne} that all operators in the
gauge theory except those in short multiplets acquire large anomalous
dimensions in the strong 't~Hooft coupling limit, and perhaps even
decouple from the operator algebra.\footnote{We thank T.~Banks for a
discussion on this point.}  The supergravity fields corresponding to
the operators in short multiplets have been tabulated in \cite{KRvN}.
Inspection of this table reveals that the only scalar $SO(6)$ singlet
operators are the renormalizable lagrangian $\O_4$ (coupling to the
$s$-wave of the dilaton) and a dimension eight operator $\O_8$ which
couples to uniform dilations of the $S^5$ part of the near-horizon
geometry.  There is also a dimension four pseudo-scalar which couples
to the axion, which we shall ignore in the following.

On the grounds of group theory and large anomalous scaling dimensions,
we are then led to the tentative conclusion that the effective
lagrangian for the low-energy dynamics at large $gN$ is
  \eqn{LGuess}{
   {\cal L} = \O_4 + R^4 \O_8  \ .
  }
 The relation to DBI is merely that at the low-energy effective
lagrangian of the same system at small $gN$ is the DBI action.  On
this view, the phrase ``DBI action'' must be interpreted in
\cite{GHKK98} (and in the many other papers in the literature,
e.g.~\cite{cgkt}, where it was invoked in the context of an effective
world-volume theory of D-brane black holes) as a metonym for its
strong-coupling relative.  \LGuess\ is a fantastic simplification over
the still incompletely known non-abelian DBI action.  But in a way it
is no less problematical as a specification of a quantum theory.  The
natural interpretation of \LGuess\ is as the Wilsonian effective
action with cutoff on the order $R$.\footnote{If the cutoff $\Lambda$
is made arbitrary, then one must introduce a coupling
$\lambda(\Lambda)$ in front of $\O_8$ which runs precisely in order to
keep the physical observables, e.g.~correlation functions, invariant
with respect to the change in the choice of the cut-off.}  The
difficulties with this approach include pinning down the normalization
of $\O_8$ at a given cutoff, defining an appropriate regularization
scheme which allows one to recover maximal supersymmetry, and the
apparent vanishing of $\langle \O_4 \O_4 \O_8 \rangle$ in the $AdS$/CFT
prescription to leading order in large $gN$.

\begin{figure}
       \centerline{\psfig{figure=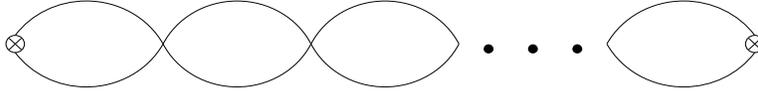,width=4in}}
   \caption{A diagram with $n$ quartic vertices contributing at 
    order $O(R^{4n})$.}
   \label{figA}
  \end{figure}

Nevertheless, let us try to argue that \LGuess\ at least has the
potential to reproduce all the correction terms in \PiXForm.
Following \cite{GHKK98}, we can consider as a toy model free $U(1)$
gauge theory with an $F^4$ interaction.  From graphs such as the one
in figure~\ref{figA},
one indeed obtains a $(R^4/x^4)^n (\log R^2/x^2)^n$ correction to the
two-point function.  It is fascinating that the final forms
(\ref{PForm}) and \PiXForm\ of the absorption probability and
two-point function are so simple and suggestive of Feynman integrals,
regulated at the scale $\mu = 1/R$.  For small $\omega R$, it seems
that the perturbative expansion around the conformal limit may be
better defined than we have any right to expect based on previous
experience with non-renormalizable divergences in quantum field
theories.  Quite remarkably, one type of interaction alone is
sufficient to reproduce the form of \PiXForm. This might indeed be a
consequence of superconformal invariance and the decoupling of
non-chiral operators in the large $gN$ limit severely restricting the
dynamics away from the infrared fixed point.  We regretfully leave a
more detailed study for future work.

\section{Discussion} 
\label{Discuss}

The biggest obstacle to finding evidence for the conjectured
throat-brane equivalence \cite{JuanAds97,gkPol,WitHolOne} between
${\cal N}=4$ super Yang-Mills theory and supergravity on $AdS_5 \times
S^5$ is that supergravity's validity is restricted to the region of
strong 't~Hooft coupling, where gauge theory calculations are
difficult.  Let us adopt units where the radius of $S^5$ is $1$.
Briefly, since $1/\alpha' \sim g_{YM} \sqrt{N}$ in these units, the
$\alpha'$ corrections to the supergravity action are important except
in the limit of large $g_{YM} \sqrt{N}$.  For example, the
supergravity fields on $AdS_5$ (with Kaluza Klein masses on the order
$1/R$) are much lighter than massive string states (with masses on the
order $1/\sqrt{\alpha'}$) only in this limit.  The corresponding
non-chiral fields in the gauge theory ``freeze out'' on account of an
anomalous dimension on the order $(g_{YM} \sqrt{N})^{1/2}$
\cite{gkPol}.  Large $N$ can be regarded as a separate requirement:
since powers of $\kappa \sim 1/N$ suppress quantum loop corrections to
supergravity, the identification of the classical supergravity action
with the generator of connected Green's functions can only capture the
leading large $N$ asymptotics.

To proceed to finite or small $g_{YM} \sqrt{N}$ seems difficult
without some profound new insight into the description of string
theory in Ramond-Ramond backgrounds.  Any hope of systematic
perturbative field theory evidence in favor of the throat-brane
conjecture would seem to depend on finding some other small coupling
parameter.  The only candidate seems to be $\omega R$, where $\omega$
is the energy of a given process (i.e.~absorption).  As a first step
in investigating a possible perturbation expansion in $\omega R$, we
have given an algorithm, which can be readily implemented on a
computer, for extracting the absorption cross-section of a minimal
scalar in an arbitrary partial wave.  The notion \cite{GHKK98,deAl}
that the DBI action of D3-branes can in any meaningful way
``holograph'' supergravity or string theory on the full extremal
three-brane geometry must be viewed with skepticism.  It is perhaps
more reasonable to hope that a quantum field theoretic derivation of
at least the leading log terms in the $\omega R$ series expansion
might be achieved (in part because these terms have a simpler cutoff
dependence than terms with fewer powers of logarithms).  In
geometrical terms, the hope would be to see the $r/R$ corrections to
the near-horizon geometry (where $r$ is the usual radial variable
entering into the harmonic function $H = 1 + R^4/r^4$) reflected order
by order in the non-renormalizable contributions to the Green's
functions for some quantum effective world volume theory.

While the motivation for this work was primarily our hope to achieve a
better understanding of the double scaling limit described in
\cite{Kleb97,GKT97}, our main technical results can be stated in the
more prosaic setting of Schrodinger operators in one dimension.  For a
particle moving in a potential $V(z) = -2q \cosh 2z$, we have found a
simple expression \AbsProb\ for the transmission coefficient in terms
of the Floquet exponent $\nu$ and a quantity $\chi$ related to the
transformation properties of Floquet solutions under parity.  The
computation of the Floquet is well understood in terms of partial
fractions.  We implement the methods of \cite{doug} to give a method
for computing $\chi$ as well.  The Hamiltonian form of \vEvolve, and
the surprising symmetry in the transmission probability between $\eta
= e^{i\pi\nu}$ and $\chi$, tantalizes us with the hope that one might
be able to give a treatment of Mathieu functions which puts $\eta$ and
$\chi$ on an equal footing.

\section*{Acknowledgements}

We would like to thank M.~Fisher for introducing us to Mathieu
functions.  Thanks to R.~Goldstein, E.~Lieb, R.~Askey, and the
participants of the String Dualities program at ITP, Santa Barbara,
for discussions.  We appreciate S. de Alwis' prompt reading of the
manuscript.  S.S.G. is grateful to the ITP for hospitality during the
initial stages of this work.  This research was supported in part by
the National Science Foundation under Grant No. PHY94-07194, by the
Department of Energy under Grant No. DE-FG02-91ER40671, and by the
James S.~McDonnell Foundation under Grant No. 91-48.  S.S.G. also
thanks the Hertz Foundation for its support.

\appendix

\section{Explicit formulas}

In this appendix we present explicit forms for some results which were
considered too lengthy to write out in the main text.  Most of the
computations were done with Mathematica.

First, the Floquet exponent for $r=1$, $r=3/2$, and $r=2$
(corresponding to $l=0$, $l=1$, and $l=2$) can be expanded as a power
series in $\lambda$ as follows:
\begin{eqnarray}
  r=1 & : & \nu = 
  2 - {\frac{i}{3}}\,{\sqrt{5}}\,{{\lambda}^4} + 
 {\frac{7\,i}{108 \sqrt{5}}}\,{{\lambda}^8} + 
   {\frac{11851\,i}{31104 \sqrt{5}}}\,{{\lambda}^{12}} + 
   \ldots \nonumber \\
r={3 \over 2} &:& \nu= 
 3 - {\frac{1}{6}}\,{{\lambda}^4} + {\frac{133}{4320}}\,{{\lambda}^8} + 
   {\frac{311}{1555200}}\,{{\lambda}^{12}} + \ldots \label{nuexp} \\
r=2 & : & \nu = 
  4 - {\frac{1}{15}}\,{{\lambda}^4} - {\frac{137}{27000}}\,{{\lambda}^8} + 
   {\frac{305843}{680400000}}\,{{\lambda}^{12}} + 
	\ldots \ . \nonumber
\end{eqnarray} 

By iterating \GotA\ one can obtain expressions for the formal series
$A_q$ defined in \DougSum\ in terms of the ``loop variables''
$S[\alpha_0,\alpha_1,\ldots,\alpha_k]$.  These grow in size very
rapidly:
\begin{eqnarray}
   A_1 &=& S[1]  \nonumber \\
   A_2 &=&  {\frac{{{S[1]}^2}}{2}} - {\frac{S[2]}{2}} - S[1,1]  \nonumber \\
   A_3 &=& {\frac{{{S[1]}^3}}{6}} - {\frac{S[1] S[2]}{2}} +
    {\frac{S[3]}{3}} - S[1] S[1,1] + 
    S[1,2] + S[2,1] + S[1,1,1]  \label{SeveralA} \\
   A_4 &=& {\frac{{{S[1]}^4}}{24}} - {\frac{{{S[1]}^2} S[2]}{4}} + 
    {\frac{{{S[2]}^2}}{8}} + {\frac{S[1] S[3]}{3}} - {\frac{S[4]}{4}}- 
    {\frac{{{S[1]}^2} S[1,1]}{2}} + {\frac{S[2] S[1,1]}{2}} +
    {\frac{{{S[1,1]}^2}}{2}} \nonumber \\
&&    +     S[1] S[1,2] - S[1,3] + S[1] S[2,1]   - {\frac{3 S[2,2]}{2}} - 
    S[3,1] + S[1] S[1,1,1] \nonumber \\
&& - S[1,1,2] - 2 S[1,2,1]       - 
    S[2,1,1] - S[1,1,1,1] \nonumber
\end{eqnarray}
and so on.  

After making the identification \SpecialT, the formal sums
$S[\{\alpha_i\}]$ may be evaluated explicitly in the manner indicated
in the paragraph following \SpecialT.
\begin{eqnarray}
S[1] & = & {\frac{-3 - 2z}
    {\left( -1 + 2r \right) \left( 1 + 2r \right) 
      \left( -1 + r - z \right) \left( 1 + r + z \right) }}
+ {\frac{\psi(1 - r + z) - \psi(1 + r + z)}{-r + 4{r^3}}}
\nonumber \\
S[2] & = &  {\frac{35 + 84z + 70{z^2} + 20{z^3} + 
      8{r^4}\left( 1 + 2z \right)  - 
      2{r^2}\left( 35 + 50z + 28{z^2} + 8{z^3} \right) }{
      {{\left( -1 + 4{r^2} \right) }^3}{{\left( 1 - r + z \right) }^2}
      {{\left( 1 + r + z \right) }^2}}}\nonumber \\
&&+{\frac{\left( -1 + 20{r^2} \right) 
      \left( \psi(1 - r + z) - \psi(1 + r + z) \right) }{2{r^3}
      {{\left( -1 + 4{r^2} \right) }^3}}}\label{ExplicitS} \\
&&+{\frac{\left( 1 + 4{r^2} \right) 
      \left( \psi^{(1)}(1 - r + z) + \psi^{(1)}(1 + r + z) \right) }{2{r^2}
      {{\left( 1 - 4{r^2} \right) }^2}}}\nonumber \\
S[1,1] & = & \nonumber \\
\lefteqn{ \!\!\!\!\!-{\frac{ 
      4{r^6}\left( 3 + 2z \right)  + 
      {r^4}\left( 35 - 26z - 36{z^2} - 8{z^3} \right)  - 
      {r^2}\left( 109 + 143z + 65{z^2} + 10{z^3} \right) +2{{\left( 2 + z \right) }^2} }{4
      \left( -1 + r \right) \left( 1 + r \right) 
      {{\left( r - 4{r^3} \right) }^2}\left( -2 + r - z \right) 
      \left( -1 + r - z \right) \left( 1 + r + z \right) 
      \left( 2 + r + z \right) }}}\nonumber \\
&& +  {\frac{\left( -1 + 10{r^2} \right) 
      \left( \psi(1 - r + z) - \psi(1 + r + z) \right) }{4{r^3}
      {{\left( 1 - 4{r^2} \right) }^2}\left( -1 + {r^2} \right) }}
+ {\frac{\psi^{(1)}(2 - r + z) + \psi^{(1)}(2 + r + z)}{4{r^2} - 16{r^4}}} \nonumber 
\end{eqnarray}
 These formulas also become very lengthy, and they have many different
forms because of the various identities for the $\psi$ function.

\bibliography{mat}
\bibliographystyle{ssg}

\end{document}